\documentclass[pre,twocolumn,amsmath,preprintnumbers,nofootinbib]{revtex4}
\usepackage{amsmath,amsfonts,bm}
\usepackage{graphicx}
\begin{document}


\noindent
{\bf Comment on ``Failure of the Work-Hamiltonian Connection for Free-Energy Calculations''}

\noindent
J.\ Horowitz and C.\ Jarzynski

\noindent
University of Maryland, College Park 20742
\vskip .1in

Nonequilibrium work relations establish a connection between nonequilibrium work values and equilibrium free energies.
In a recent Letter, Vilar and Rubi~\cite{Vilar08} (VR) argue that the definition of work used in these relations,
\begin{equation}
\label{eq:W}
W = \int dt \, \frac{\partial H}{\partial t} = \int dt \, \dot\lambda \, \frac{\partial H}{\partial\lambda} ,
\end{equation}
is incorrect, and therefore the relations themselves are fundamentally flawed.
In our investigations, however, we have reached the opposite conclusion~\cite{JH}, as have Imparato and Peliti~\cite{Imparato07} in direct response to VR.

In Eq.~\ref{eq:W}, $\lambda$ represents generalized coordinates ($a_1, a_2,\cdots$) describing the external bodies that we manipulate to act on the system of interest; $W$ is the integral of force, multiplied by the displacements of these bodies~\cite{Gibbs_Sekimoto}.
Referring to Refs.~\cite{JH,Imparato07,Gibbs_Sekimoto} for a broader discussion, in this Comment we illustrate that $W$ has exactly the properties we associate with thermodynamic work.

\begin{figure}[htbp] 
   \centering
   \includegraphics[width=2in,angle=-90]{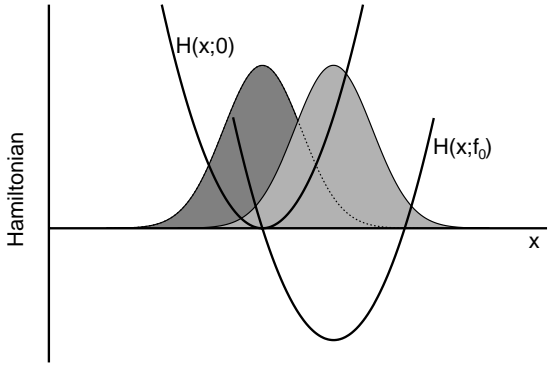}
   \caption{The Hamiltonian functions and equilibrium distributions (shaded) at $f=0$ and $f=f_0$.}
   \label{fig:H}
\end{figure}

We consider the model system analyzed in Ref.~\cite{Vilar08}, described by the Hamiltonian
\begin{equation}
\label{eq:H}
H(x;f) = \frac{1}{2} k x^2 - fx ,
\end{equation}
and evolving under Langevin dynamics with friction coefficient $\gamma=1$.
The force $f(t)$ is switched on uniformly, from $f(0)=0$ to $f(\tau)=f_0$;
outside the time interval $0<t<\tau$ the force is held constant.
This system evolves from equilibrium state $A$ in the distant past ($f=0$), to equilibrium state $B$ in the distant future ($f=f_0$).
The initial and final Hamiltonian functions and canonical distributions are shown in Fig.~\ref{fig:H}.
Since $H(x;f(t))$ is {\it constant} for $t<0$ and $t>\tau$, its value during these times is identified with the energy of the system~\cite{Vilar08}.
Using the equilibrium distribution $p \propto \exp(-\beta H)$,  we compute the internal energy ($E = \int p H$) and the entropy ($S = -\int p\ln p$) for states $A$ and $B$:
$E_A = (2\beta)^{-1}$, $E_B = (2\beta)^{-1} - f_0^2/(2k)$, $S_A=S_B=[1-\ln(\beta k/2\pi)]/2$.
From the {\it thermodynamic} definition of free energy, $G = E - ST$~\cite{Landau}, we then get
\begin{equation}
\label{eq:dG}
\Delta G = G_B - G_A = - \frac{f_0^2}{2k} .
\end{equation}
The negative value of $\Delta G$ reflects a decrease in internal energy, with no change in entropy (see Fig.~\ref{fig:H}).

For this model, the distribution of work values over an ensemble of realizations of the process, $\rho(W)$, can be obtained using the approach of Ref.~\cite{Mazonka99}.
This distribution is a Gaussian with mean and variance,
\begin{equation}
\label{eq:gaussian}
\langle W \rangle =  - \frac{f_0^2}{2k} + \frac{\beta \sigma_W^2}{2}
\quad , \quad
\sigma_W^2 = \frac{2f_0^2}{\beta k^2\tau}
\left[ 1 + \frac{e^{-k\tau}-1}{k\tau}\right].
\end{equation}
This result implies that:
$W\rightarrow\Delta G$ for every realization in the reversible limit ($\tau\rightarrow\infty$);
$\langle W\rangle > \Delta G$ in the irreversible case (finite $\tau$); and
$\langle e^{-\beta W}\rangle = e^{-\beta\Delta G}$ for any value of $\tau$.
Thus {\it the work defined by Eq.~\ref{eq:W} is consistent with the second law of thermodynamics, and its exponential average correctly gives $\Delta G$}, when the free energy is defined by the expression $G = E-ST$. 

By contrast, VR obtain $\Delta G>0$ for this model (Eq.~4 of Ref.~\cite{Vilar08}), and assert that a negative value would be ``inconsistent with a nonspontaneous process''.
We disagree.
An {\it undisturbed} system certainly seeks to minimize its free energy (e.g., after the removal of a constraint), but
when an external agent varies a parameter of the system, such as the field $f$ above, then {\it there is no universal restriction on the sign of the free energy change}.
For instance, by manipulating a piston we can either increase or decrease the Helmholtz free energy of a gas, according to whether we compress or expand the gas.


\begin{thebibliography}{99}
\bibitem{Landau}
The distinction between Gibbs and Helmoltz free energies is not relevant for this model.
\end{thebibliography}
\end{document}